\documentclass[aps, prl, twocolumn, floatfix, superscriptaddress]{revtex4-2}
\usepackage{amsmath, amssymb, graphicx, epsfig, xcolor}
\usepackage[colorlinks=true, allcolors=blue]{hyperref}

\begin{document}

\title {PBE-GGA Predicts the B8$\leftrightarrow$B2 Phase Boundary of FeO at Earth's Core Conditions}

\author{Zhen Zhang}
\address{Department of Applied Physics and Applied Mathematics, Columbia University, New York, NY 10027, USA}

\author{Yang Sun}
\address{Department of Applied Physics and Applied Mathematics, Columbia University, New York, NY 10027, USA}

\author{Renata M. Wentzcovitch}
\address{Department of Applied Physics and Applied Mathematics, Columbia University, New York, NY 10027, USA}
\address{Department of Earth and Environmental Sciences, Columbia University, New York, NY 10027, USA}
\address{Lamont-Doherty Earth Observatory, Columbia University, Palisades, NY 10964, USA}

\date{October 21, 2022}

\begin{abstract}
FeO is a crucial phase of the Earth's core, and its thermodynamic properties are essential to developing more accurate core models. It is also a notorious correlated insulator in the NaCl-type (B1) phase at ambient conditions. It undergoes two polymorphic transitions at 300 K before it becomes metallic in the NiAs-type (B8) structure at $\sim$100 GPa. Although its phase diagram is not fully mapped, it is well established that the B8 phase transforms to the CsCl-type (B2) phase at core pressures and temperatures. Here we report a successful \textit{ab initio} calculation of the B8$\leftrightarrow$B2 phase boundary in FeO at Earth's core pressures. We show that fully anharmonic free energies computed with the PBE-GGA + Mermin functional reproduce the experimental phase boundary within uncertainties at $P > 240$ GPa, including the largely negative Clapeyron slope of $-52 \pm 5$ MPa/K. This study validates the applicability of a standard DFT functional to FeO under Earth's core conditions and demonstrates the theoretical framework that enables complex predictive studies of this region.
\end{abstract}

\maketitle

FeO is one of the major constituents of Earth and other terrestrial planets. It is not only the iron end-member of ferropericlase ((Mg$_{1-x}$Fe$_{x}$)O), the second most abundant phase in the Earth's lower mantle, but also a significant alloying component in the Earth's core \cite{ref1}. As a classic ``correlated'' oxide, phase relations in FeO are also of great interest in condensed matter physics. Like ferropericlase, FeO undergoes a spin state change under pressure \cite{ref2, ref3}. FeO exhibits rich phenomenology at high pressures and temperatures ($P,T$), e.g., polymorphic, magnetic, and insulating-metallic transitions. Such phase changes control fundamental material properties of the Earth, e.g., its thermal and electrical conductivity and magnetic susceptibility, to mention a few.

FeO has an enigmatic phase diagram (e.g., see \cite{ref4}). It is stable in the cubic NaCl-type (B1) structure at ambient conditions. Under compression at room temperature, B1 undergoes a phase transition to a rhombohedral B1 (rB1) structure above 16 GPa \cite{ref5}. While rB1 is believed to be the ground state \cite{ref6} at low temperatures and low pressures, B1 remains stable at higher temperatures along the geotherm throughout the mantle \cite{ref2, ref7, ref8, ref9}. A further change to the NiAs-type (B8) structure was observed above 90 GPa at 600 K \cite{ref10}. The B1$\leftrightarrow$rB1 and rB1$\leftrightarrow$B8 phase boundaries have been measured at and above 300 K \cite{ref10, ref11}. A B1$\leftrightarrow$B8 phase boundary has also been measured at high temperatures up to $\sim$240 GPa, a typical outer core pressure \cite{ref2, ref7, ref8, ref9}. In addition, the monoclinic B1 \cite{ref12, ref13} and the inverse B8 phases \cite{ref11, ref14} have also been observed at low temperatures.

Besides these five crystal structures, FeO's CsCl-type (B2) structure \cite{ref15, ref16} has been identified at higher $P,T$. Unlike MgO, CaO, SrO, and BaO, which transform directly from the B1 to the B2 structure, FeO has the intermediate and partially covalent/metallic B8 phase \cite{ref4, ref10} up to $\sim$3800 K and $\sim$240 GPa \cite{ref15}. The direct B1$\leftrightarrow$B2 transformation in FeO occurs only above this temperature \cite{ref15}, with a B1, B2, and B8 triple point expected near these conditions. So far, only one experimental study has reported measurements of B2-related phase boundaries \cite{ref15}. Electronic structure and spin states of the B1 \cite{ref2, ref3, ref8, ref17, ref18, ref19, ref20} and B8 \cite{ref20, ref21, ref22, ref23, ref24} phases have often been investigated, given the theoretical challenge of dealing with the interplay between electronic correlation and structural phase transitions. 

Here we perform \textit{ab initio} calculations of the B8$\leftrightarrow$B2 boundary at $\sim$150--400 GPa, a relevant pressure range for the Earth's core, and $T > 1000$ K. Under such extreme conditions, anharmonicity is fundamental in determining dynamic and thermodynamic stabilities, especially for the B2 phase (e.g., the analogous \textit{bcc} phase of elemental metals \cite{ref25, ref26, ref27, ref28, ref29} and the B2 phase of binary compounds \cite{ref30}). The high-spin (HS) to low-spin (LS) and insulator-to-metal transitions in FeO happen at $\sim$120 GPa \cite{ref24, ref31}, a typical mantle pressure. A recent density functional plus dynamical mean-field theory (DFT+DMFT) study \cite{ref20} confirmed the LS and metallic (delocalized) electronic state in both B2 and B8 phases under core conditions. These observations suggest that DFT-based molecular dynamics (DFT-MD) could address the B8$\leftrightarrow$B2 phase competition. 

\begin{figure}[t]
	\includegraphics[width=\linewidth]{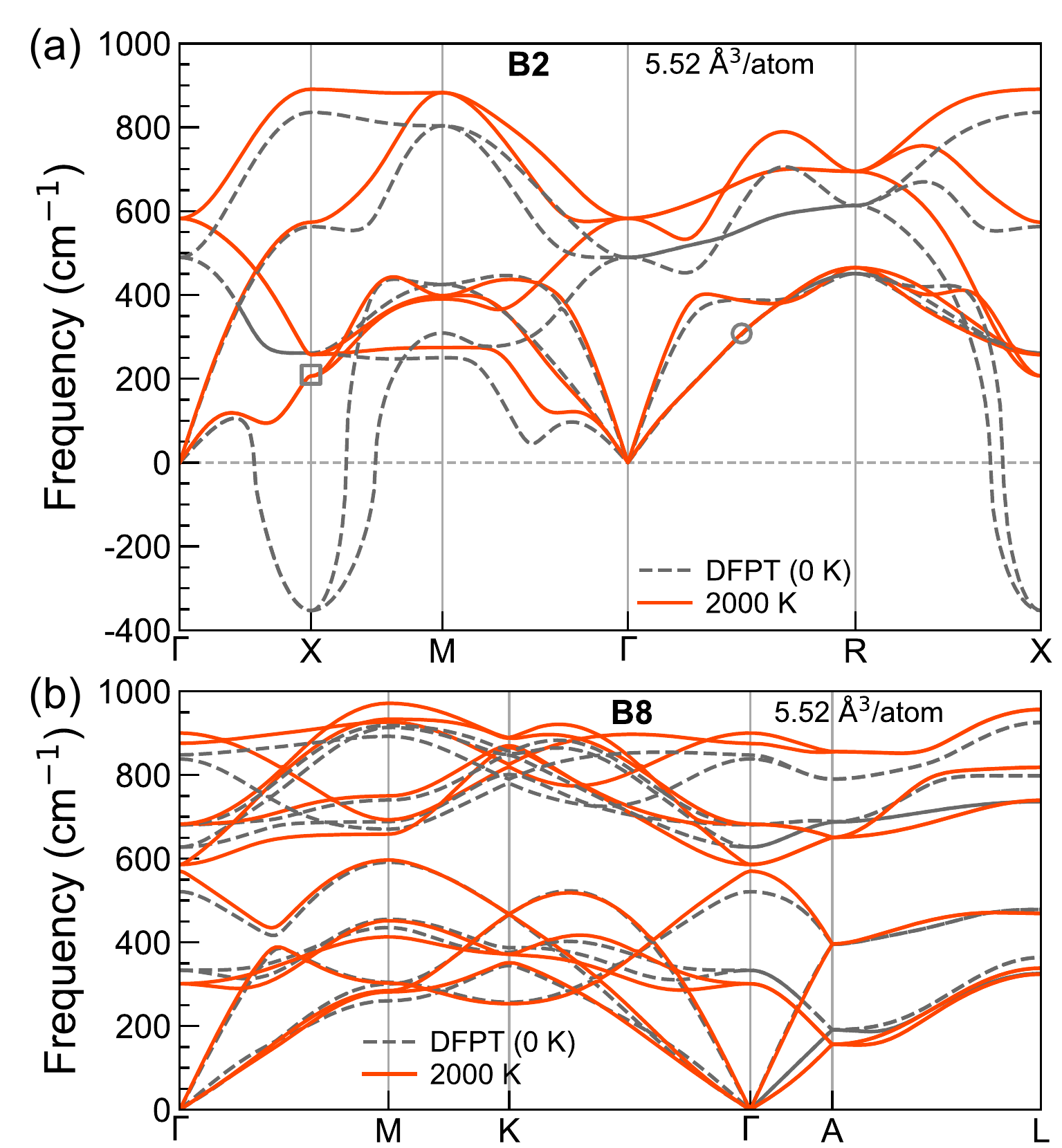}
	\caption{Harmonic (dashed gray curves) and anharmonic phonon dispersions at 2000 K (solid orange curves) for (a) B2 and (b) B8 phases at $V = 5.52$ $\mathrm{\AA^3}$/atom.}
	\label{fig1}
\end{figure}

We first address harmonic phonon dispersions at 0 K in both B2 and B8 phases using density-functional perturbation theory (DFPT) \cite{ref32}. DFT and DFPT calculations were performed using the PAW method \cite{ref33} as implemented in Quantum ESPRESSO \cite{ref34}. We used the Perdew-Burke-Ernzerhof (PBE) \cite{ref35} generalized gradient approximation (GGA) to compute the exchange-correlation energy. Phonons were calculated on $4\times4\times4$ and $4\times4\times2$ \textbf{q}-meshes in the B2 and B8 structures, respectively. Figures. \ref{fig1}(a) and \ref{fig1}(b) show harmonic phonon dispersions (dashed gray curves) for these phases. The B2 phase displays unstable modes around phonon wave vector $\mathbf{q}=\mathbf{X}\left(0, \frac{1}{2}, 0\right)$, while the B8 phase's phonons are all stable. The harmonic phonon instability in the B2 phase is not out of expectation. A similar soft mode behavior in the \textit{bcc} phase of many elemental metals \cite{ref25, ref26, ref27, ref28, ref29} drives the \textit{bcc} to \textit{hcp} phase transition at low temperatures. The B2 and B8 phases of FeO are analogous to the elemental \textit{bcc} and \textit{hcp} phases, respectively (see Fig. S1 of the Supplemental Materials (SM) \cite{ref36}). Note that though B8 FeO shares the same $P6_3/mmc$ space group with the elemental \textit{hcp} phase, the atomic layers stacking in B8 ($A^{\mathrm{Fe}} B^{\mathrm{O}} A^{\mathrm{Fe}} C^{\mathrm{O}} \ A^{\mathrm{Fe}} B^{\mathrm{O}} A^{\mathrm{Fe}} C^{\mathrm{O}}$) differs from that in elemental \textit{hcp} ($AB \ AB$). Imaginary frequencies at 0 K indicate that phonon-phonon interactions are critical in the dynamic stabilization of this structure, and entropic effects at high temperatures can stabilize it thermodynamically. Conventional harmonic phonons or quasiharmonic free energies cannot address the lattice dynamics and thermodynamic properties of this B2 phase.

We employ the phonon quasiparticle (PHQ) approach \cite{ref37, ref38} to address the high-temperature anharmonicity of FeO under core $P,T$ conditions. This approach assumes that a system with fully interacting phonons can be simplified as an effective system with noninteracting PHQs \cite{ref39, ref40}. Each PHQ can be described by two parameters, renormalized frequency, $\widetilde{\omega}_{\mathbf{q} s}$, and linewidth, $\Gamma_{\mathbf{q} s}$. A PHQ is numerically defined by mode-projected velocity autocorrelation function (VAF) \cite{ref37, ref38},
\begin{equation}
	\left\langle V_{\mathbf{q} s}(0) V_{\mathbf{q} s}(t)\right\rangle=\lim _{\tau \rightarrow \infty} \frac{1}{\tau} \int_0^\tau V_{\mathbf{q} s}^*\left(t^{\prime}\right) V_{\mathbf{q} s}\left(t^{\prime}+t\right) d t^{\prime},
	\label{eq1}
\end{equation}
where $V_{\mathbf{q} s}(t)=\sum_{i=1}^N \sqrt{M_i} \mathbf{v}_i(t) e^{i \mathbf{q} \cdot \mathbf{R}_i} \cdot \hat{\mathbf{e}}_{\mathbf{q} s}^i$ is the mass-weighted and $\mathbf{q} s$-mode-projected velocity. $M_i$, $\mathbf{R}_i$, and $\mathbf{v}_i$ $(i=1,\dots,N)$ are atomic mass, crystallographic atomic coordinate, and atomic velocity computed by DFT-MD of an \textit{N}-atom supercell. $\hat{\mathbf{e}}_{\mathbf{q} s}^i$ is the harmonic phonon eigenvector obtained at an electronic temperature $T_\mathrm{el}=T$ in the DFT-MD \cite{ref29, ref39, ref40}, where \textbf{q} should be commensurate with the supercell size and $s$ labels the phonon branches at each \textbf{q}. For a well-defined PHQ, its VAF assumes an exponentially decaying cosine form $A_{\mathbf{q} s} \cos \left(\widetilde{\omega}_{\mathbf{q} s} t\right) e^{-\Gamma_{\mathbf{q} s} t}$, where $A_{\mathbf{q} s}$ is the initial oscillation amplitude. The well-defined VAF's power spectrum $\left|\int_0^{\infty}\left\langle V_{\mathbf{q} s}(0) V_{\mathbf{q} s}(t)\right\rangle e^{i \omega t} d t\right|^2$ assumes a Lorentzian line shape with a single peak at $\widetilde{\omega}_{\mathbf{q} s}$ and a linewidth $\Gamma_{\mathbf{q} s}$.

To compute PHQs, \textit{ab initio} molecular dynamics (AIMD) simulations were performed using the PAW PBE as implemented in VASP \cite{ref41}. The electronic temperature ($T_\mathrm{el}$) was set the same as the ionic temperature ($T$) using the Mermin functional \cite{ref42, ref43}. FeO was simulated with 128-atom supercells ($4\times4\times4$ for B2 and $4\times4\times2$ for B8) with a $\Gamma$ \textbf{k}-point sampling and a kinetic energy cutoff of 400 eV. The supercells are sufficiently large to converge the harmonic part of interatomic force constants. Thus they should also be sufficiently large to converge the anharmonic part since the anharmonic part of interatomic force constants has shorter ranges than the harmonic one \cite{ref38, ref44}. Simulations were conducted in the $NVT$ ensemble on a series of volumes between 4.92 and 5.88 $\mathrm{\AA^3}$/atom and temperatures between 1000 and 4000 K controlled by Nos{\'e} thermostat \cite{ref45, ref46}. Each simulation ran for 50 ps, sufficiently long to converge PHQ parameters, with a time step of 1 fs.

Well-defined PHQs obtained on the AIMD-sampled \textbf{q}-mesh enable calculations of renormalized, i.e., anharmonic phonon dispersion using Fourier interpolation \cite{ref37, ref38}. Figures \ref{fig1}(a) and \ref{fig1}(b) show the anharmonic phonon dispersions obtained at $T = 2000$ K (solid orange curves) for the B2 and B8 phases, respectively. The unstable mode with imaginary harmonic frequency at $\mathbf{q}=\mathbf{X}\left(0, \frac{1}{2}, 0\right)$ in the B2 phase stiffens drastically at high temperatures (see the gray square in Fig. \ref{fig1}(a) and Fig. S2 of the SM \cite{ref36}). The anharmonic phonon dispersions are free of imaginary frequencies, indicating that B2 FeO is stabilized dynamically by phonon-phonon interactions at high temperatures \cite{ref29, ref38, ref47}. Frequency renormalization in the B8 phase is not as significant as in the B2, yet also not negligible. Thus, the following free energy calculations use temperature-dependent anharmonic phonon spectra (see Fig. S3 of the SM \cite{ref36}) for both phases. The anharmonic phonon spectra were evaluated on much denser \textbf{q}-meshes ($20\times20\times20$ for B2 and $20\times20\times10$ for B8) via Fourier interpolation \cite{ref37, ref38} to approximate the thermodynamic limit.

\begin{figure}[t]
	\includegraphics[width=\linewidth]{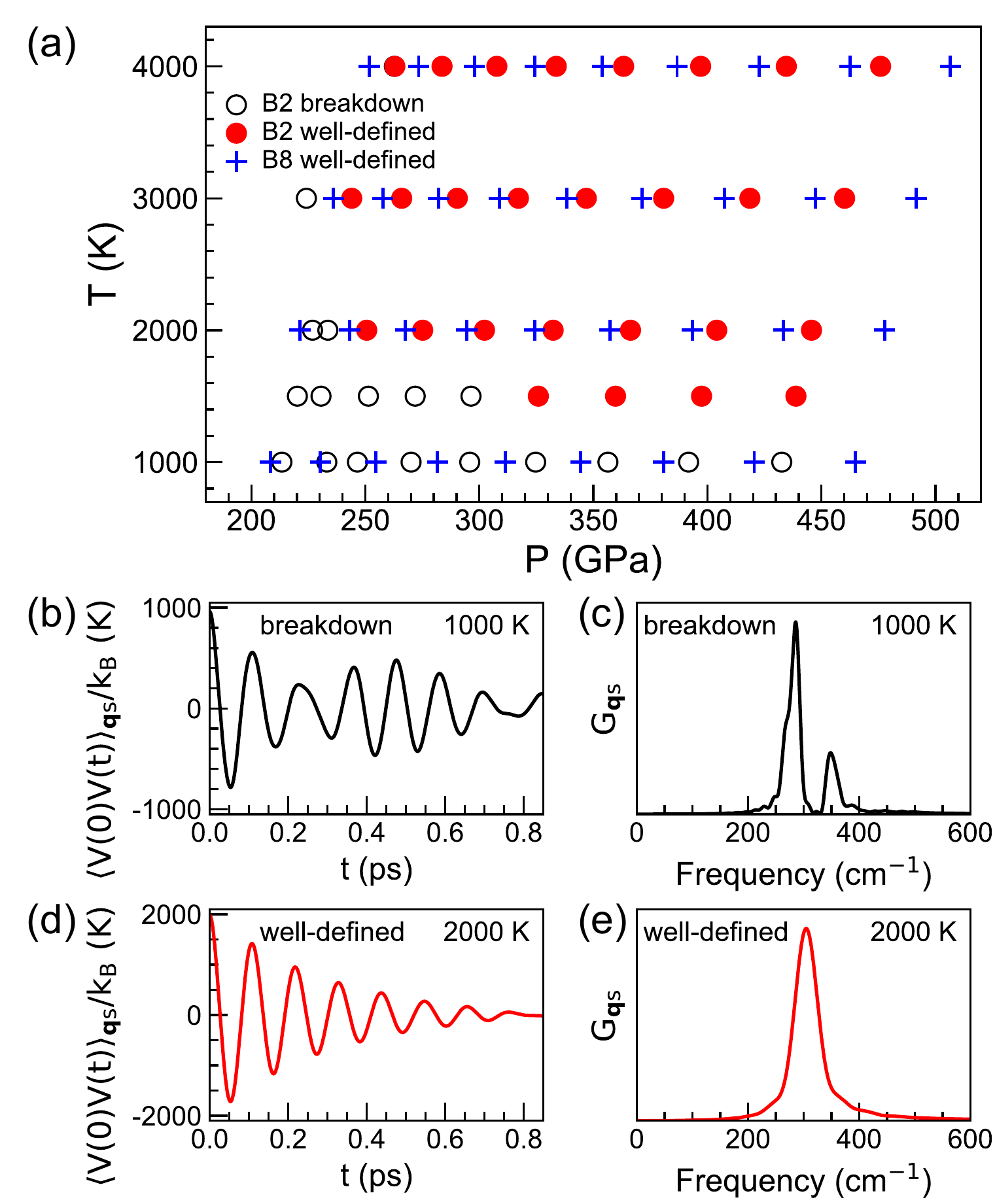}
	\caption{(a) $P,T$ conditions covered by AIMD simulations. Hollow circles indicate conditions where the B2 structure is unstable and/or PHQs are not well-defined. Red circles indicate a dynamically stable B2 phase with well-defined PHQs. Blue crosses indicate a dynamically stable B8 phase. Pressures are ensemble averages from AIMD simulations. Mode-projected VAF of a TA mode at $\mathbf{q}=\left(\frac{1}{4}, \frac{1}{4}, \frac{1}{4}\right)$ (gray circle in Fig. \ref{fig1}(a)) with a harmonic frequency of 304 cm$^{-1}$ for B2 at $V = 5.52$ $\mathrm{\AA^3}$/atom and $T =$ (b) 1000 and (d) 2000 K, respectively. (c) and (e) show the corresponding power spectra.}
	\label{fig2}
\end{figure}

We performed AIMD simulations for both B8 and B2 phases at the $P,T$ conditions indicated in Fig. \ref{fig2}(a). The corresponding $V,T$ conditions are shown in Fig. S4 of the SM \cite{ref36}. The B8 phase is dynamically stable at all studied $P,T$s, while the B2 phase shows structural and/or phononic instabilities at certain low $P,T$s. For instance, at $T = 1000$ K and $P =$ $\sim$213 GPa ($V = 5.88$ $\mathrm{\AA^3}$/atom), the eight-fold coordinated Fe in the starting B2 structure transforms to a six-fold coordinated structure after thermal equilibration (see Fig. S5 of the SM \cite{ref36}). By removing the B2 lattice constraints, a complete phase transition to B8 is realized. Compressed from $\sim$213 to $\sim$270 GPa ($V = 5.52$ $\mathrm{\AA^3}$/atom) at $T = 1000$ K, the structure no longer shows the B2$\rightarrow$B8 transition. PHQs in the B2 phase are still not well-defined at this $P,T$. Fig. \ref{fig2}(b) shows the 1000 K VAF for the transverse acoustic mode at $\mathbf{q}=\left(\frac{1}{4}, \frac{1}{4}, \frac{1}{4}\right)$ (gray circle in Fig. \ref{fig1}(a)), which exhibits a pattern far distinct from an exponentially decaying cosine form. The corresponding power spectrum shows two peaks (Fig. \ref{fig2}(c)), indicating the breakdown of a well-defined B2 phase PHQ \cite{ref29, ref38}. This behavior signals the tendency of atoms to displace from the B2 equilibrium sites and the B2 structure to distort. In contrast, this mode is stable at 2000 K ($\sim$275 GPa at the same $V = 5.52$ $\mathrm{\AA^3}$/atom), as indicated by the well-defined PHQ with an exponentially decaying VAF (Fig. \ref{fig2}(d)) and a single Lorentzian-shaped peak in the power spectrum (Fig. \ref{fig2}(e)). As seen in Fig. \ref{fig2}(a), higher $P,T$s systematically stabilize the B2 phase.

The extensive AIMD results indicated in Fig. \ref{fig2}(a) enable Gibbs free energy calculations on an equal footing for the B2 and B8 phases in a large $P,T$ range. With Fourier interpolated anharmonic phonon spectra, the vibrational entropy can be obtained in the thermodynamic limit within the phonon gas model (PGM) \cite{ref37, ref38, ref48}, 
\begin{equation}
	S_{\mathrm{vib}}(T)=k_{\mathrm{B}} \sum_{\mathbf{q} s}\left[\left(n_{\mathbf{q} s}+1\right) \ln \left(n_{\mathbf{q} s}+1\right)-n_{\mathbf{q} s} \ln n_{\mathbf{q} s}\right],
	\label{eq2}
\end{equation}
where $n_{\mathbf{q} s}=\left[\exp \left(\hbar \widetilde{\omega}_{\mathbf{q} s}(T) / k_{\mathrm{B}} T\right)-1\right]^{-1}$. $\widetilde{\omega}_{\mathbf{q} s}(T)$ were obtained by fitting calculated $\widetilde{\omega}_{\mathbf{q} s}$'s at several temperatures and constant volume to a second-order polynomial in $T$ \cite{ref29, ref37, ref47}. The Helmholtz free energy at constant volume can be obtained by integrating both the electronic and vibrational entropies \cite{ref29, ref47},
\begin{eqnarray}
	F(T)=E\left(T_0\right)-T_0 \left[S_{\mathrm{el}}\left(T_0\right)+S_{\mathrm{vib}}\left(T_0\right)\right] \nonumber\\
	-\int_{T_0}^T \left[S_{\mathrm{el}}\left(T^{\prime}\right)+S_{\mathrm{vib}}\left(T^{\prime}\right)\right] d T^{\prime},
	\label{eq3}
\end{eqnarray}
where $T_0$ is a reference temperature at which all PHQs are well-defined, $E(T_0)$ and $S_{\mathrm{el}} (T_0)$ are time-averaged internal (potential + kinetic) energy and electronic entropy obtained from AIMD at $T_0$. The choice of $T_0$ does not change the resulting thermodynamics, so we set $T_0 = 4000$ K. $S_{\mathrm{el}} (T)$ at constant volume was computed as the ensemble average at temperatures shown in Figs. \ref{fig2}(a) and S4 of the SM \cite{ref36} and fit to a second-order polynomial in $T$ \cite{ref29}. Contrary to $S_{\mathrm{vib}}$ that suffers from finite-size effects and requires Fourier interpolation, $S_{\mathrm{el}}$ and $E$ are more insensitive to the simulation cell size \cite{ref40}, and AIMD ensemble averages converge well.

\begin{figure}[t]
	\includegraphics[width=\linewidth]{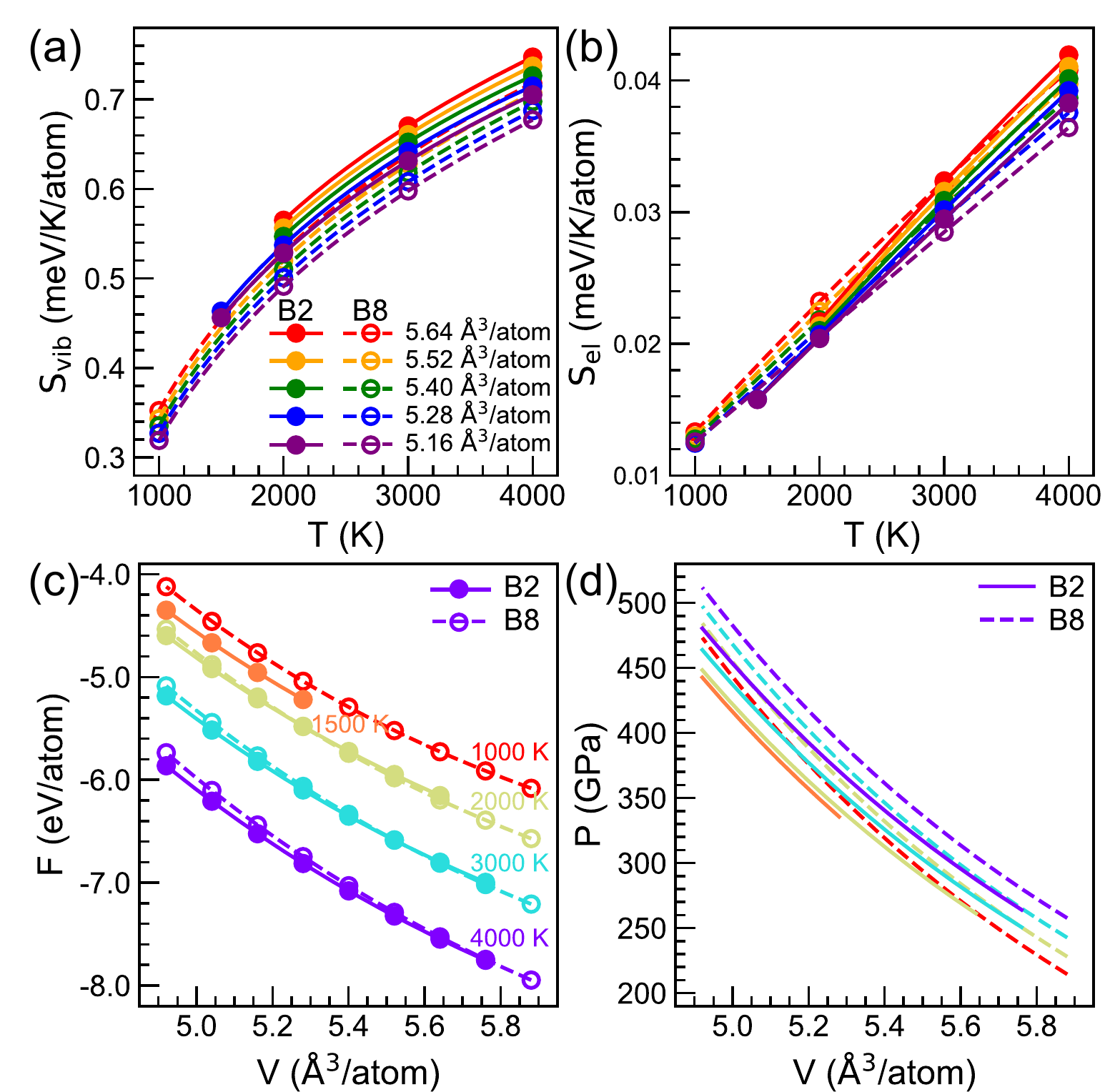}
	\caption{(a) Vibrational entropy $S_{\mathrm{vib}} (V,T)$ and (b) electronic entropy $S_{\mathrm{el}} (V,T)$ versus $T$ at different $V$s. (c) Helmholtz free energy $F(V,T)$ and (d) pressure $P(V,T)$ versus $V$ at different $T$s. B2 is shown by solid circles and solid curves, and B8 is shown by open circles and dashed curves. Circles indicate the $V,T$s sampled by the AIMD simulations.}
	\label{fig3}
\end{figure}

At the same $V,T$ conditions, $S_{\mathrm{vib}}$ is always larger for B2 than for B8 (Fig. \ref{fig3}(a)). The generally lower renormalized frequencies in B2 (see Fig. S3 of the SM \cite{ref36}) contribute to its larger entropy and thermodynamic stability with respect to B8 at higher temperatures. $S_{\mathrm{el}}$ varies nearly linearly with temperature (Fig. \ref{fig3}(b)), and a quadratic fitting for $S_{\mathrm{el}}(T)$ is sufficiently accurate \cite{ref29}. $S_{\mathrm{el}}$'s of both phases are similar but much smaller than $S_{\mathrm{vib}}$ (see Figs. \ref{fig3}(a) and \ref{fig3}(b)). Therefore, $S_{\mathrm{vib}}$ dominates entropic effects on the free energy. Calculated $F(V,T)$s are shown in Fig. \ref{fig3}(c). Isothermal equations of state (EOS) were computed by fitting $F(V)$ to a third-order finite strain expansion. Pressures calculated as $P=-\left(\frac{\partial F}{\partial V}\right)_T$ are shown in Fig. \ref{fig3}(d). At the same $P,T$ conditions, the stable B2 phase volume is always smaller than the B8 volume, and the difference $\Delta(P V)=(P V)_{\mathrm{B} 8}-(P V)_{\mathrm{B} 2}$ increases with pressure (see Fig. S6 of the SM \cite{ref36}), which contribute to the enthalpic stabilization of B2 at high pressures. We thereby predict a negative Clapeyron slope ($\frac{d P}{d T}=\frac{\Delta S}{\Delta V}$) for the phase boundary.

\begin{figure}[t]
	\includegraphics[width=\linewidth]{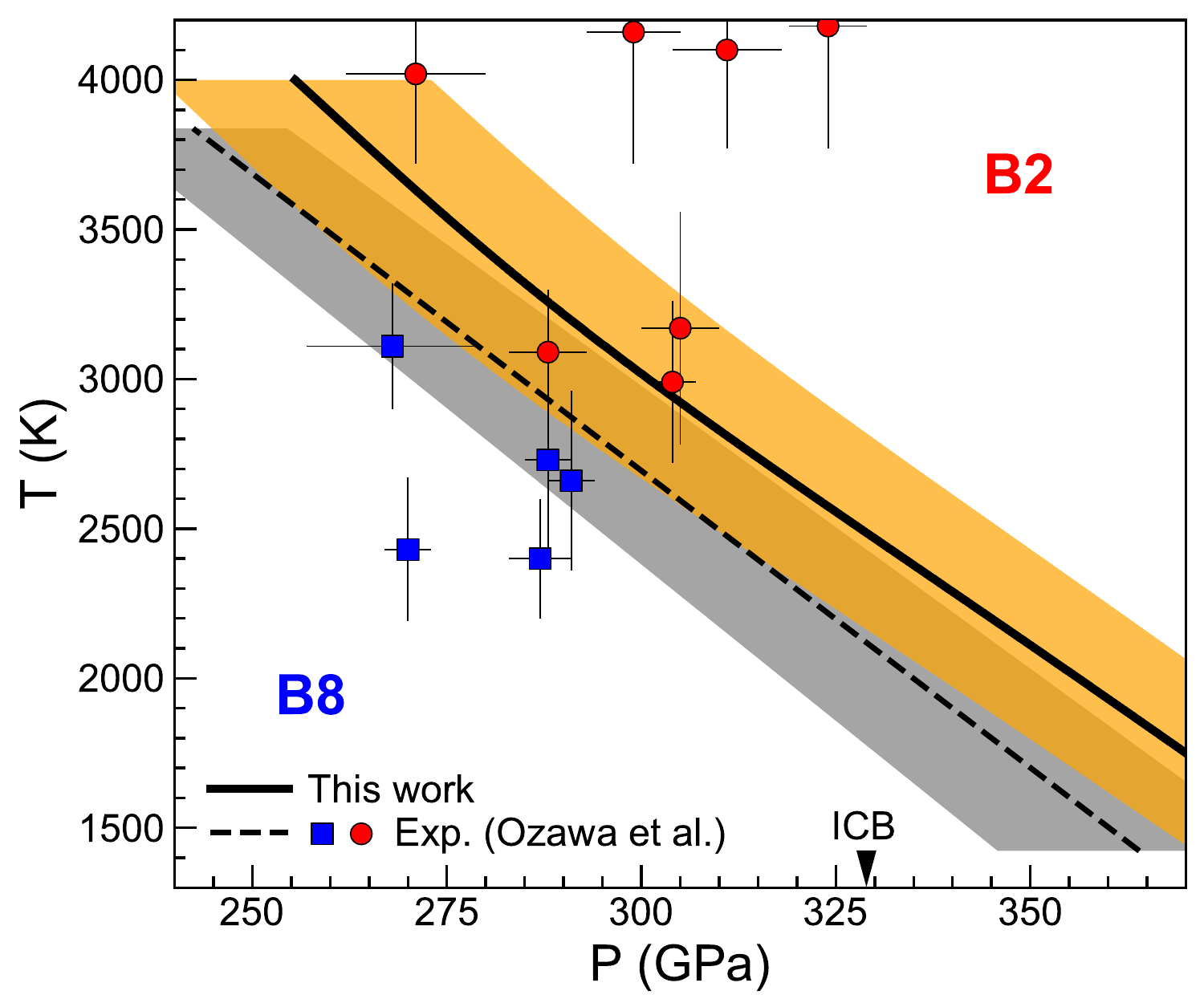}
	\caption{B8$\leftrightarrow$B2 phase boundary of FeO. The solid black curve is the calculated phase boundary, and the shaded orange area indicates the computational uncertainty. The x-ray diffraction measurements \cite{ref15} for B2 (red circles with error bars) and B8 (blue squares with error bars) phases are shown for comparison. The dashed line is the experimental phase boundary \cite{ref15}, and the shaded gray area indicates a recent estimation of its uncertainty \cite{ref52}. The arrow indicates the inner core boundary (ICB) pressure.}
	\label{fig4}
\end{figure}

The Gibbs free energy was calculated as $G(P,T)=F(V,T)+P(V,T)V$ utilizing the fitted EOS. Comparing the Gibbs free energies of both phases (see Fig. S7 of the SM \cite{ref36}), we obtain the B8$\leftrightarrow$B2 phase boundary shown in Fig. \ref{fig4}. The uncertainty in our free energy and phase boundary calculations are estimated as follows: Fig. S8 of the SM \cite{ref36} shows the dependence of the Helmholtz free energy difference, $\Delta F$, between the two phases on the \textbf{k}-mesh sampling. We consider the results on a much more computationally expensive $3\times3\times3$ \textbf{k}-mesh in the 128-atom supercell to be fully converged since it differs by less than 1 meV/atom from the $2\times2\times2$ \textbf{k}-mesh result. Hence, the difference between the $\Gamma$-point and $3\times3\times3$ \textbf{k}-mesh results is $\sim$9 meV/atom. This is the adopted uncertainty in $\Delta F$ arising from \textbf{k}-mesh sampling. The uncertainty in $\Delta F$ arising from fluctuations in AIMD simulations is estimated by conducting five parallel runs at a constant volume and 4000 K, which gives $\sim$0.1 meV/atom. Note that the dense \textbf{q}-mesh sampled in the entropy calculation (Eq. (\ref{eq2})) mimics a sufficiently large supercell (16000 atoms) calculation in the thermodynamic limit. These combined effects give an uncertainty in $\Delta F$ of $\sim$9 meV/atom, and this value is passed to $\Delta G$. This estimated uncertainty in the free energy difference compares well with the $\sim$10 meV/atom uncertainty reported in the calculation of the melting curve of iron \cite{ref39} using the same PHQ approach. It is also similar to uncertainties in other free energy difference calculations using thermodynamic integration \cite{ref49}. The uncertainty in the $PV$ term given by fitting the isothermal EOS at several volumes is a second-order effect, thus was disregarded here. The free energy uncertainty leads to an uncertainty of $\pm$18 GPa in the transition pressure, shown as the shaded orange area in Fig. \ref{fig4}. The accuracy of our prediction, however, is better than the precision. The difference between the predicted and measured \cite{ref15} phase boundaries is $\sim$5\% of the pressure, which is a typical uncertainty by \textit{ab initio} calculations (e.g., see \cite{ref50}). The error bars in Fig. \ref{fig4} label the reported experimental uncertainties in $P,T$ \cite{ref15}. A recent estimation of the uncertainty in the experimental transition pressure \cite{ref15} resulting from the choice of the Fe EOS \cite{ref51} as a pressure scale is also $\sim$5\% of the pressure \cite{ref52} and is shown as the shaded gray area in Fig. \ref{fig4}. As such, the level of agreement between our predictions and measurements of this phase boundary is excellent.

In the temperature range shown, the B8$\leftrightarrow$B2 transition occurs at $P >$ $\sim$240 GPa. The melting properties of FeO are beyond the scope of this study. At the inner-core boundary (ICB) pressure, 329 GPa \cite{ref15}, the calculated transition temperature is 2490 K. Under such conditions, the B8$\rightarrow$B2 transition is accompanied by a 1.5\% density increase. Like the analogous \textit{hcp}$\leftrightarrow$\textit{bcc} transition in elemental metals \cite{ref28, ref40, ref53}, the Clapeyron slope of the B8$\leftrightarrow$B2 transition is also negative, $-52 \pm 5$ MPa/K, which is in excellent agreement with that measured by experiments, $-50$ MPa/K \cite{ref15}.

In summary, we have investigated the B8$\leftrightarrow$B2 phase boundary of FeO at high $P,T$ conditions of the Earth's core with \textit{ab initio} calculations. We computed anharmonic free energies in the thermodynamic limit using phonon quasiparticle dispersions. The calculated phase boundary agrees with experimental observations \cite{ref15} within uncertainties. The successful calculation of the B8$\leftrightarrow$B2 phase boundary demonstrates that the PBE-GGA functional describes well energy differences between the B8 and B2 phases of FeO at these $P,T$ conditions. We might attribute this success to two factors: 1) FeO is metallic and non-magnetic at the relevant $P,T$s. A comparison between the electronic density of states calculated with PBE-GGA and DFT+DMFT \cite{ref20} (see Fig. S9 of the SM \cite{ref36}) shows qualitatively similar electronic structures near the Fermi level; 2) at the relevant $P,T$s, anharmonic effects on the B2 phase are dominant, and the vibrational entropy differences between the two phases are also well described by PBE-GGA. Therefore, the present results establish a theoretical framework for future predictive studies of the Fe-FeO system at these high $P,T$ conditions, which is key to understanding the debated problem of oxygen partitioning between the liquid and the solid regions of the Earth's core and their density deficits.

\begin{acknowledgments}
This work was primarily funded by the US Department of Energy Grant DE-SC0019759 and partly by the National Science Foundation (NSF) Grant EAR-1918126. This work used the Extreme Science and Engineering Discovery Environment, USA, supported by NSF Grant ACI-1548562. Computations were performed on Stampede2, the flagship supercomputer at the Texas Advanced Computing Center, the University of Texas at Austin, funded by NSF Grant ACI-1134872.
\end{acknowledgments}

\bibliography{feo}

\end{document}

\title {Supplemental Material for:\\PBE-GGA Predicts the B8$\leftrightarrow$B2 Phase Boundary of FeO at Earth's Core Conditions}

\author{Zhen Zhang}
\address{Department of Applied Physics and Applied Mathematics, Columbia University, New York, NY 10027, USA}

\author{Yang Sun}
\address{Department of Applied Physics and Applied Mathematics, Columbia University, New York, NY 10027, USA}

\author{Renata M. Wentzcovitch}
\address{Department of Applied Physics and Applied Mathematics, Columbia University, New York, NY 10027, USA}
\address{Department of Earth and Environmental Sciences, Columbia University, New York, NY 10027, USA}
\address{Lamont-Doherty Earth Observatory, Columbia University, Palisades, NY 10964, USA}

\maketitle

\renewcommand{\thefigure}{S\arabic{figure}}
\renewcommand{\theequation}{S\arabic{equation}}

\begin{figure}[t]
	\includegraphics[width=\linewidth]{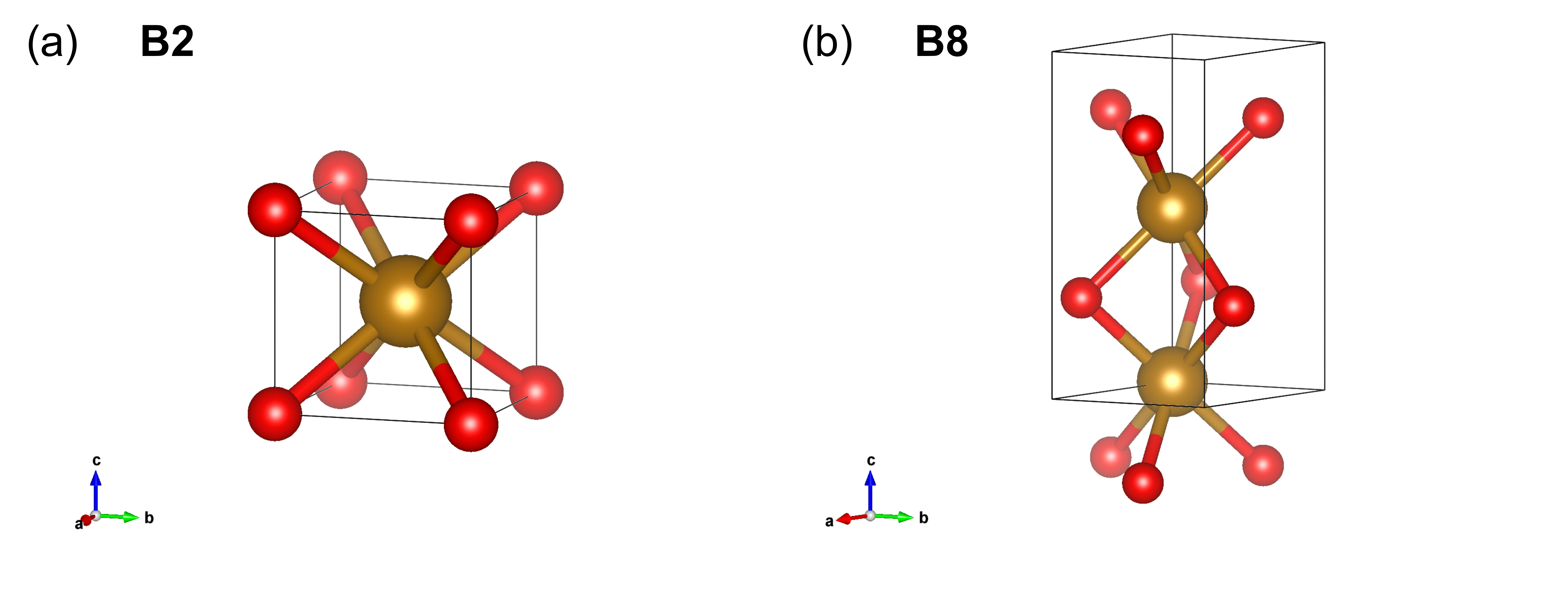}
	\caption{The crystal structure of (a) CsCl-type B2 phase and (b) NiAs-type B8 phase of FeO. Iron is shown in golden, and oxygen in red. B2 has a 2-atom primitive cell in the cubic lattice, and B8 has a 4-atom primitive cell in the hexagonal lattice.}
	\label{figs1}
\end{figure}

\begin{figure}[t]
	\includegraphics[width=\linewidth]{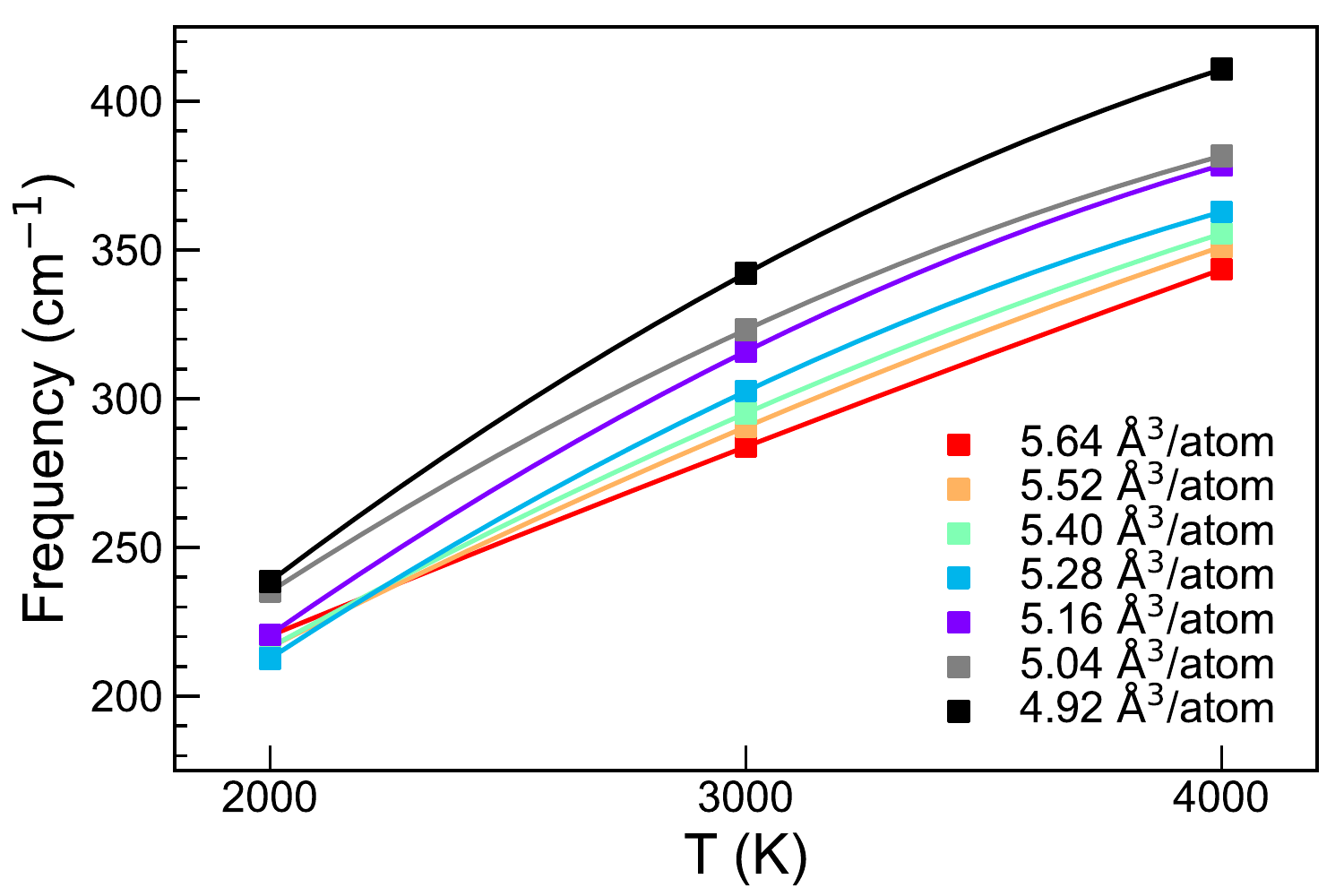}
	\caption{Temperature-dependent renormalized frequency of the transverse phonon mode at $\mathbf{q}$ = $\mathbf{X}(0,\frac{1}{2},0)$ (gray square in Fig. 1(a)) with imaginary harmonic frequency for B2 calculated at constant volume. All soft modes at different volumes acquire real renormalized frequencies and stiffen drastically with increasing temperature.}
	\label{figs2}
\end{figure}

\begin{figure}[t]
	\includegraphics[width=\linewidth]{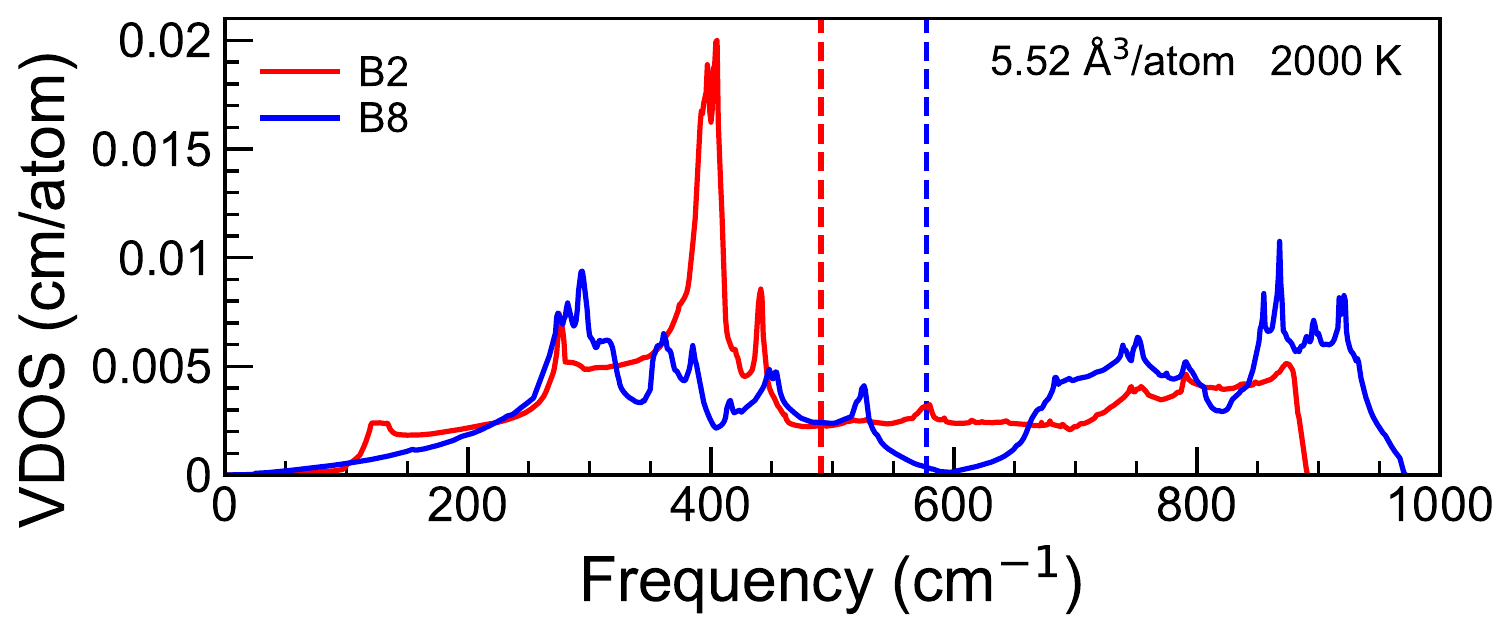}
	\caption{Anharmonic vibrational density of states (VDOS) for B2 (red curve) and B8 (blue curve) at 2000 K and $V$ = 5.52 \AA$^3$/atom. The vertical dashed lines indicate the average renormalized frequencies.}
	\label{figs3}
\end{figure}

\begin{figure}[t]
	\includegraphics[width=\linewidth]{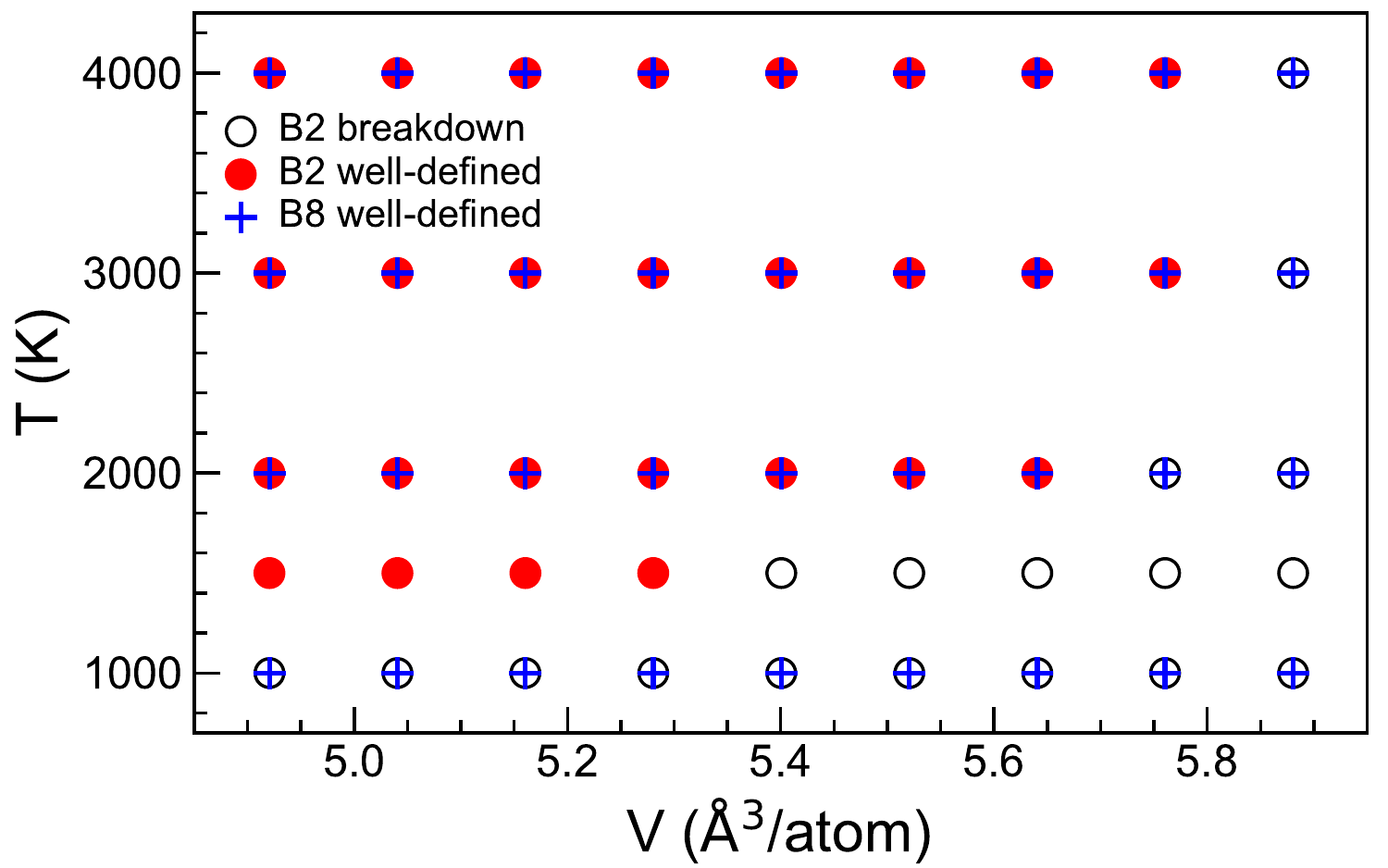}
	\caption{$V,T$ conditions covered in the AIMD simulations. Conditions at which structures and/or phonon quasiparticles for B2 break down are indicated by hollow circles. Conditions at which both structures and phonon quasiparticles are well-defined are indicated by red circles and blue crosses for B2 and B8, respectively.}
	\label{figs4}
\end{figure}

\begin{figure}[t]
	\includegraphics[width=\linewidth]{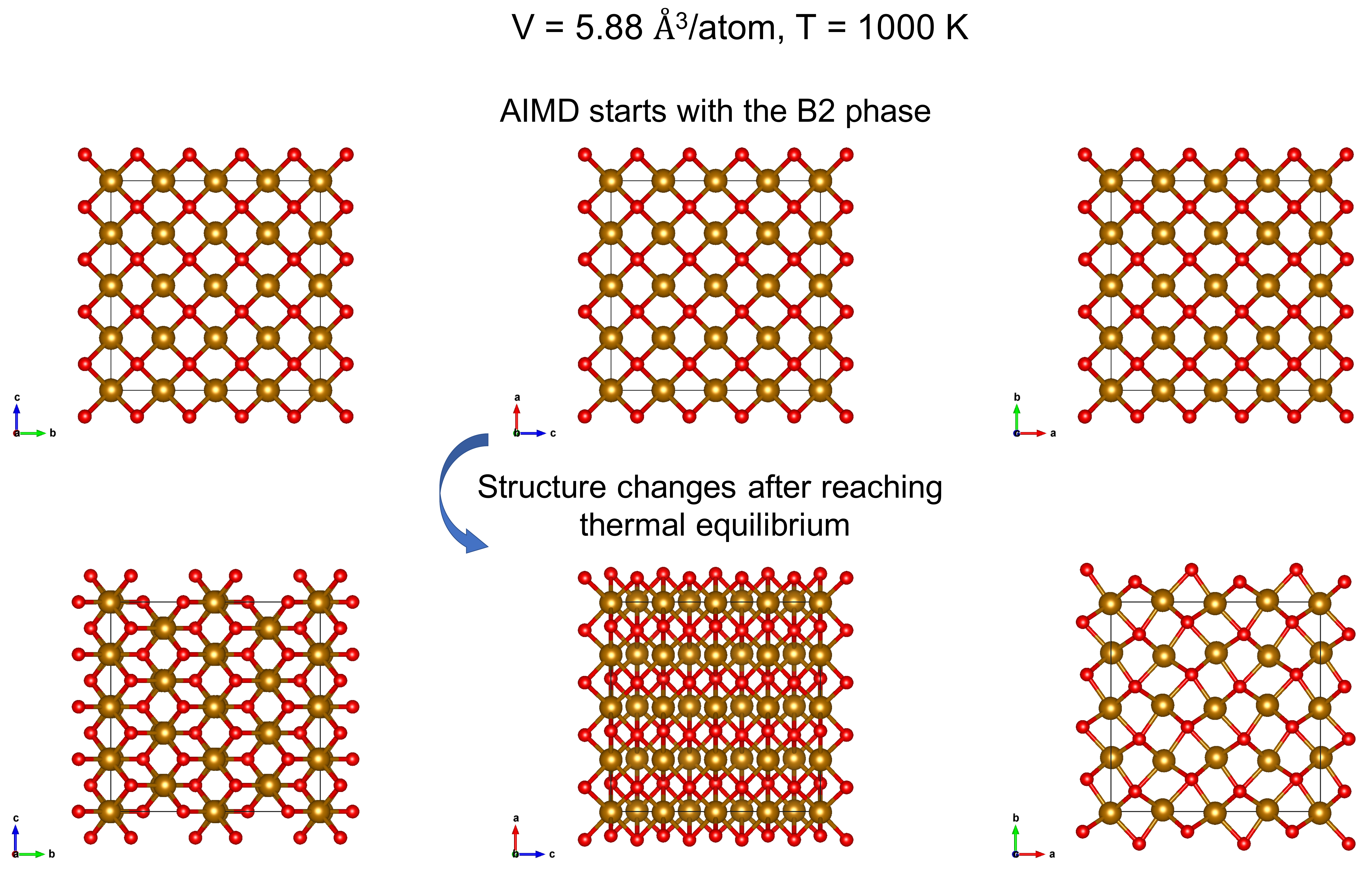}
	\caption{Time-averaged supercell structure in the B2 lattice after reaching thermal equilibrium in the AIMD simulated at $V$ = 5.88 \AA$^3$/atom and $T$ = 1000 K. The simulation starts with the B2 phase, but the B2 structure is mechanically unstable at this condition. Iron is shown in golden, and oxygen in red.}
	\label{figs5}
\end{figure}

\begin{figure}[t]
	\includegraphics[width=0.6\linewidth]{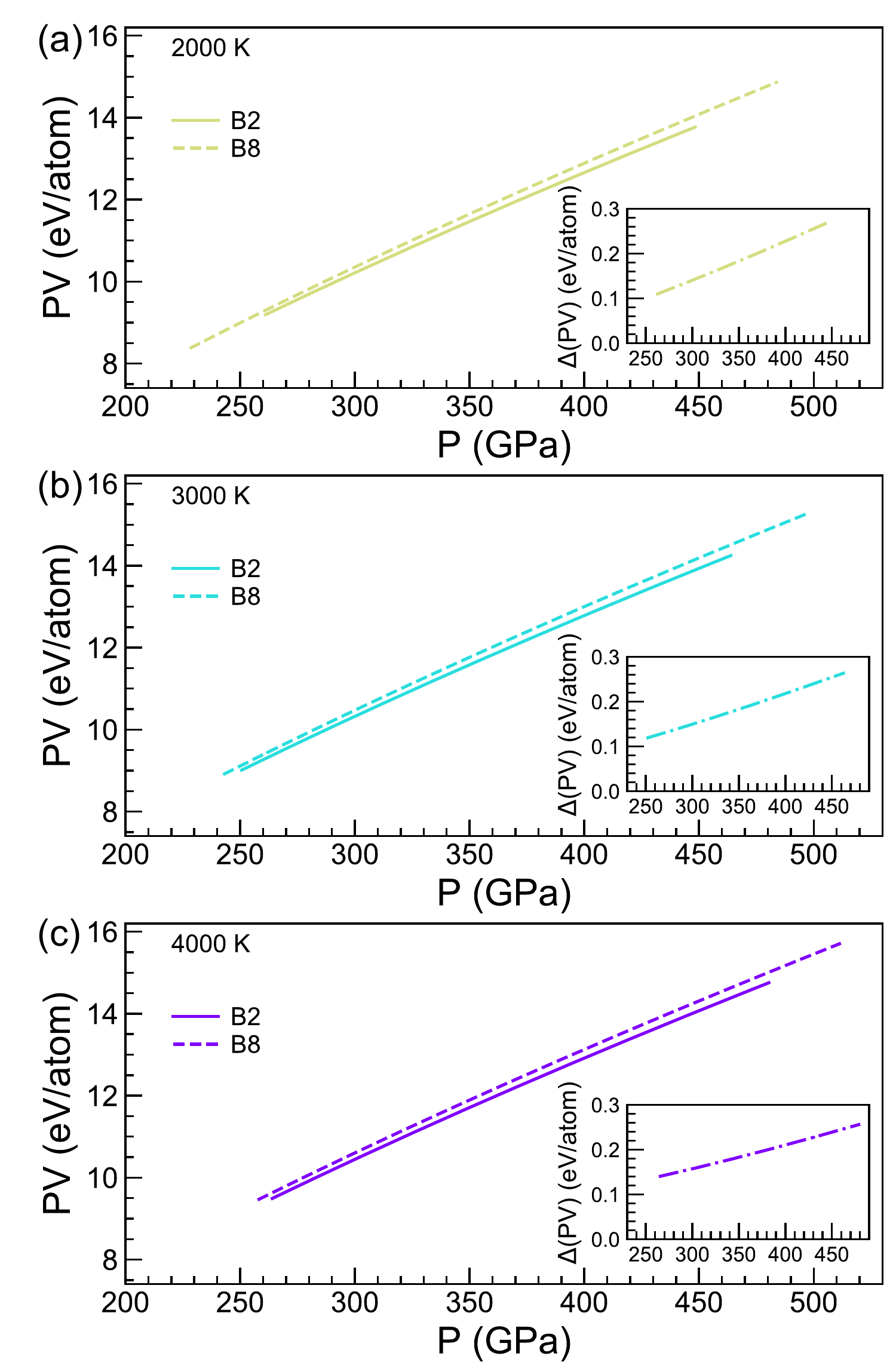}
	\caption{$PV$ versus pressure for B2 (solid curves) and B8 (dashed curves) at (a) 2000, (b) 3000, and (c) 4000 K, respectively. Inserts: $\Delta(P V)=(P V)_{\mathrm{B} 8}-(P V)_{\mathrm{B} 2}$ as a function of pressure at each temperature.}
	\label{figs6}
\end{figure}

\begin{figure}[t]
	\includegraphics[width=\linewidth]{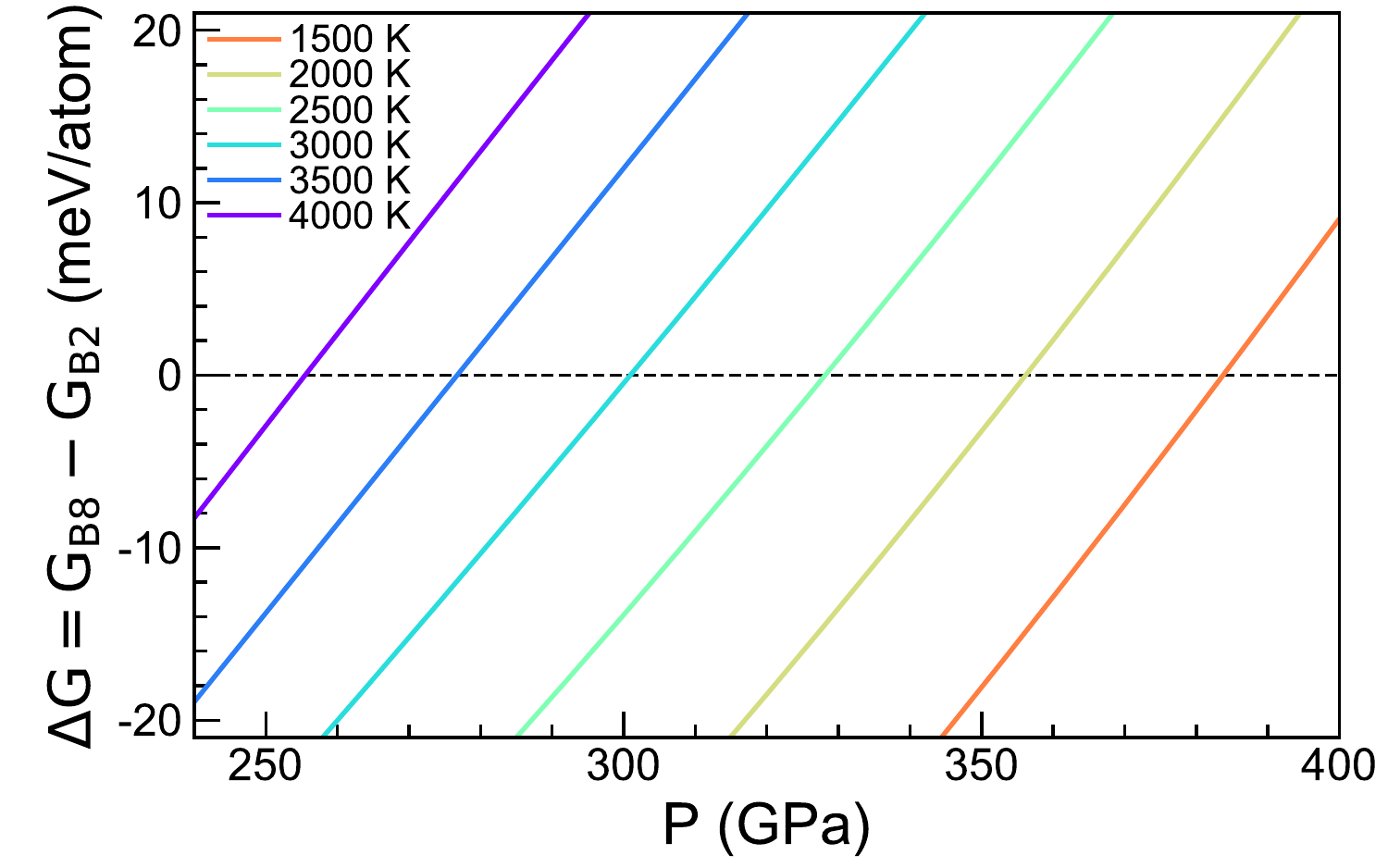}
	\caption{Gibbs free energy difference $\Delta G=G_{\mathrm{B} 8}-G_{\mathrm{B} 2}$ as a function of pressure at different temperatures.}
	\label{figs7}
\end{figure}

\begin{figure}[t]
	\includegraphics[width=0.6\linewidth]{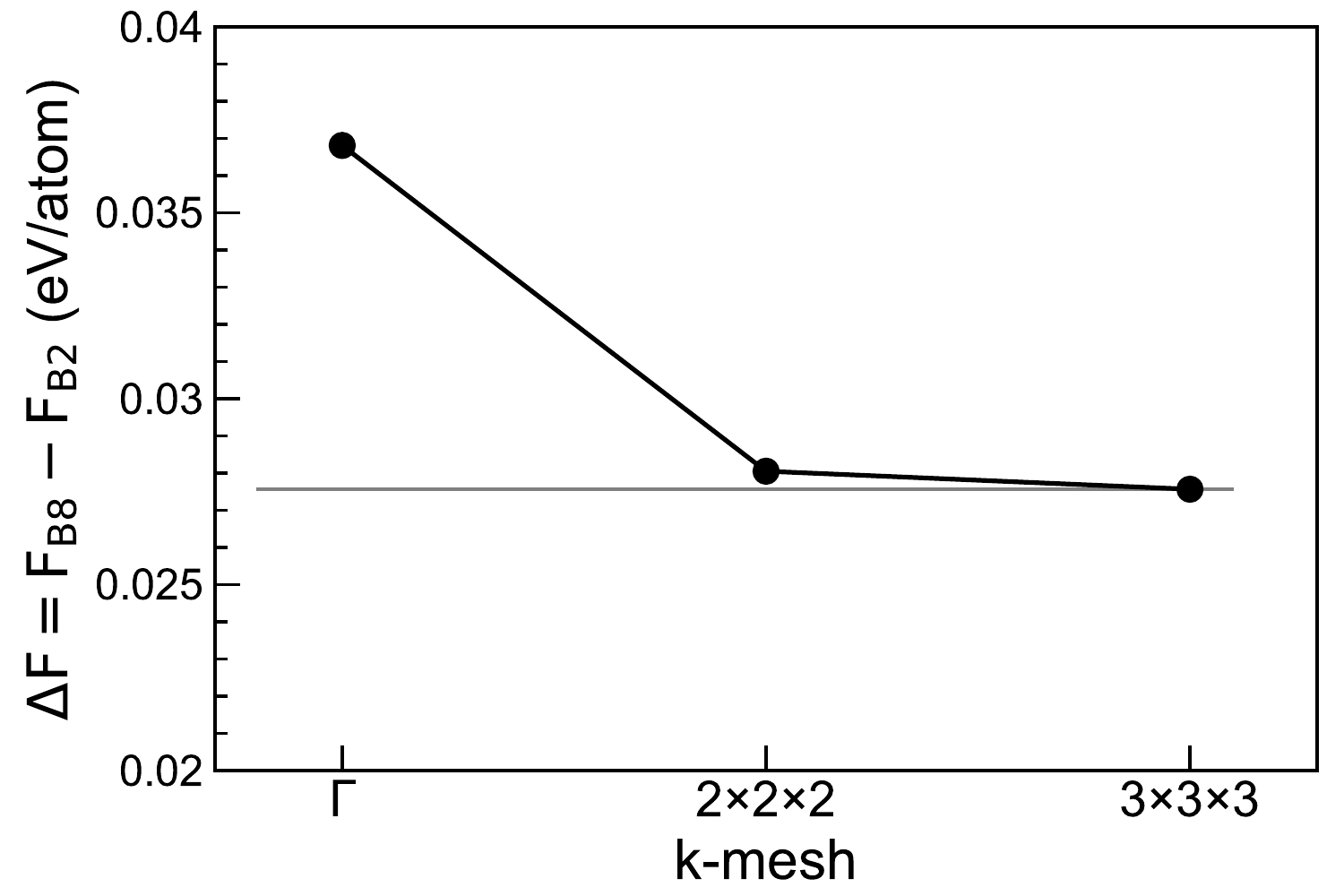}
	\caption{Helmholtz free energy difference $\Delta F=F_{\mathrm{B} 8}-F_{\mathrm{B} 2}$ calculated with $\Gamma$, $2\times2\times2$, and $3\times3\times3$ $\mathbf{k}$-mesh sampling for the 128-atom supercell at $V$ = 5.52 \AA$^3$/atom and $T$ = 4000 K. The computational error in $\mathbf{k}$-mesh sampling is $\sim$9 meV/atom by comparing $\Delta F$ between $\Gamma$ and $3\times3\times3$ $\mathbf{k}$-mesh sampling. The $\Delta F$ variance between $2\times2\times2$ and $3\times3\times3$ $\mathbf{k}$-mesh sampling is $<$1 meV/atom.}
	\label{figs8}
\end{figure}

\begin{figure}[t]
	\includegraphics[width=\linewidth]{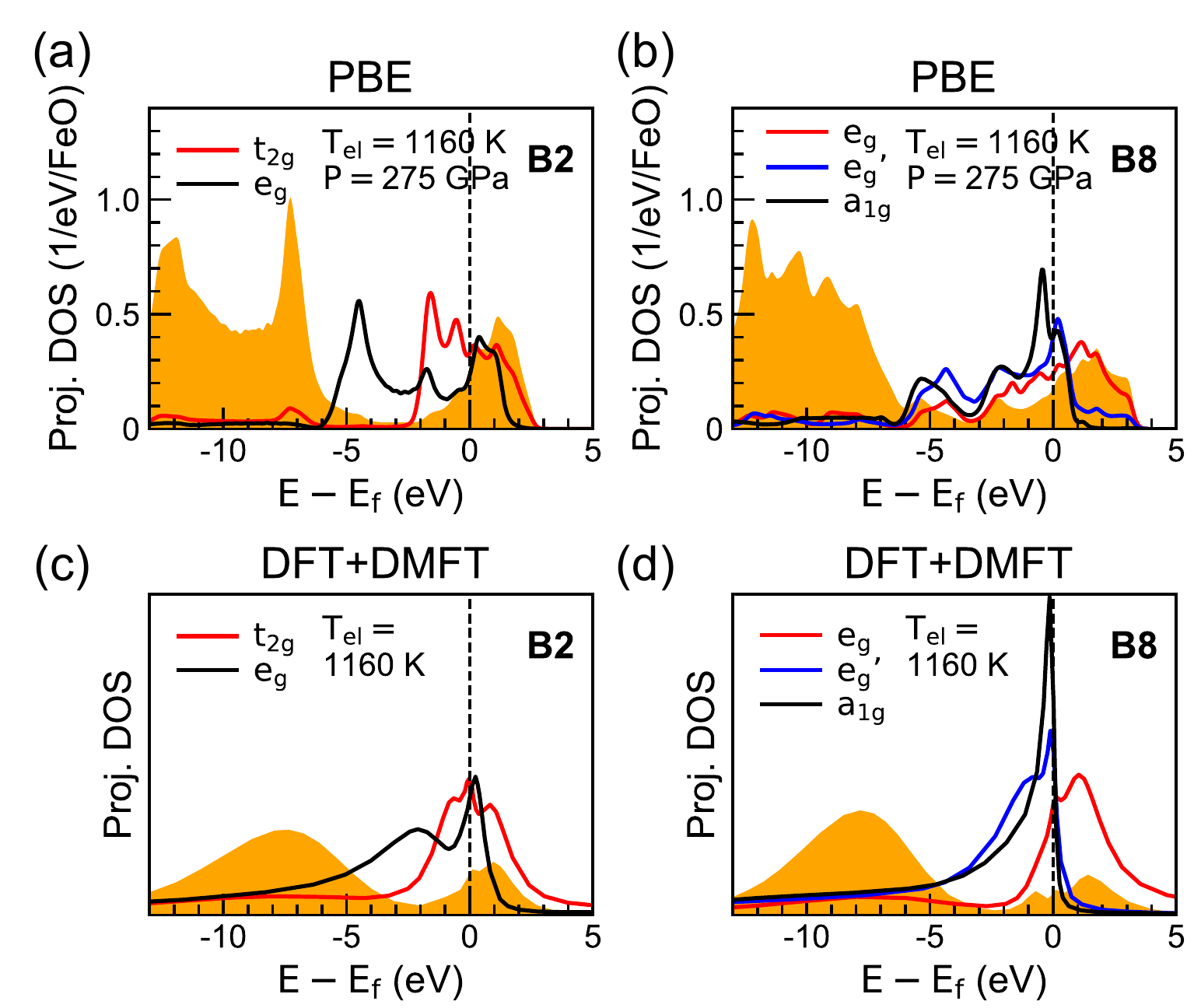}
	\caption{Projected electronic density of states (DOS) for (a) B2 and (b) B8 FeO calculated by the PBE-GGA at $T_{\mathrm{el}}$ = 1160 K and static $P$ = 275 GPa. Contributions from Fe $3d$ states are shown by solid curves, and those from O $2p$ states are shown by shaded orange areas. Projected electronic DOS for (c) B2 and (d) B8 FeO calculated by the DFT+DMFT \footnote{E. Greenberg, R. Nazarov, A. Landa, J. Ying, R. Q. Hood, B. Hen, R. Jeanloz, V. B. Prakapenka, V. V. Struzhkin, G. K. Rozenberg, and I. Leonov, Phase transitions and spin-state of iron in FeO at the conditions of earth's deep interior, arXiv:2004.00652.} at $T_{\mathrm{el}}$ = 1160 K and relevant pressures are exhibited for comparison.}
	\label{figs9}
\end{figure}

\begin{center}
{\bf Content}
\end{center}
{\bf FIG. \ref{figs1} to \ref{figs9}}

\appendix

